\documentclass[sigconf]{acmart}
\usepackage{amsmath}
\usepackage{algorithmic}
\usepackage{graphicx}
\usepackage{textcomp}
\usepackage{booktabs}
\usepackage{multirow}
\usepackage{subfigure}
\usepackage{xcolor}
\usepackage{fancyhdr}
\fancypagestyle{noheader}{
  \fancyhf{}
}

\usepackage{hyperref}
\usepackage{comment}
\DeclareMathOperator{\sgn}{sgn}
\AtBeginDocument{%
  \providecommand\BibTeX{{%
    \normalfont B\kern-0.5em{\scshape i\kern-0.25em b}\kern-0.8em\TeX}}}

\copyrightyear{2022} 
\acmYear{2022} 
\setcopyright{acmlicensed}
\acmConference[ICMI '22 Companion]{INTERNATIONAL CONFERENCE ON MULTIMODAL INTERACTION}{November 7--11, 2022}{Bengaluru, India}
\acmBooktitle{INTERNATIONAL CONFERENCE ON MULTIMODAL INTERACTION (ICMI '22 Companion), November 7--11, 2022, Bengaluru, India}
\acmPrice{15.00}
\acmDOI{10.1145/3536220.3563685}
\acmISBN{978-1-4503-9389-8/22/11}

\begin{document}

\sloppy
\title{To Improve Is to Change: Towards Improving Mood Prediction by Learning Changes in Emotion}
\renewcommand{\shorttitle}{Towards Improving Mood Prediction by Learning Changes in Emotion}


\author{Soujanya Narayana}
\affiliation{%
 \institution{University of Canberra}
 \state{ACT}
 \country{Australia}}
 \email{soujanya.narayana@canberra.edu.au}

\author{Ramanathan Subramanian}
\affiliation{%
  \institution{University of Canberra}
  \state{ACT}
  \country{Australia}}
  \email{ram.subramanian@canberra.edu.au}
  
\author{Ibrahim Radwan}
\affiliation{%
  \institution{University of Canberra}
  \state{ACT}
  \country{Australia}}
\email{ibrahim.radwan@canberra.edu.au}

\author{Roland Goecke}
\affiliation{%
  \institution{University of Canberra}
  \state{ACT}
  \country{Australia}}
\email{roland.goecke@ieee.org}
\renewcommand{\shortauthors}{S. Narayana, R. Subramanian, I. Radwan, R. Goecke}


\begin{abstract}
Although the terms \emph{mood} and \emph{emotion} are closely related and often used interchangeably, they are distinguished based on their duration, intensity and attribution.  To date, hardly any computational models have (a) examined mood recognition, and (b) modelled the interplay between mood and emotional state in their analysis. In this paper, as a first step towards mood prediction, we propose a framework that utilises both dominant emotion (or \textit{mood}) labels, and emotional change labels on the AFEW-VA database. Experiments evaluating unimodal (trained only using mood labels) and multimodal (trained with both mood and emotion change labels) convolutional neural networks confirm that incorporating emotional change information in the network training process can significantly improve the mood prediction performance, thus highlighting the importance of modelling emotion and mood simultaneously for improved performance in affective state recognition. 
\end{abstract}



\begin{teaserfigure}
\centering
  \includegraphics[height=4cm, width=\textwidth]{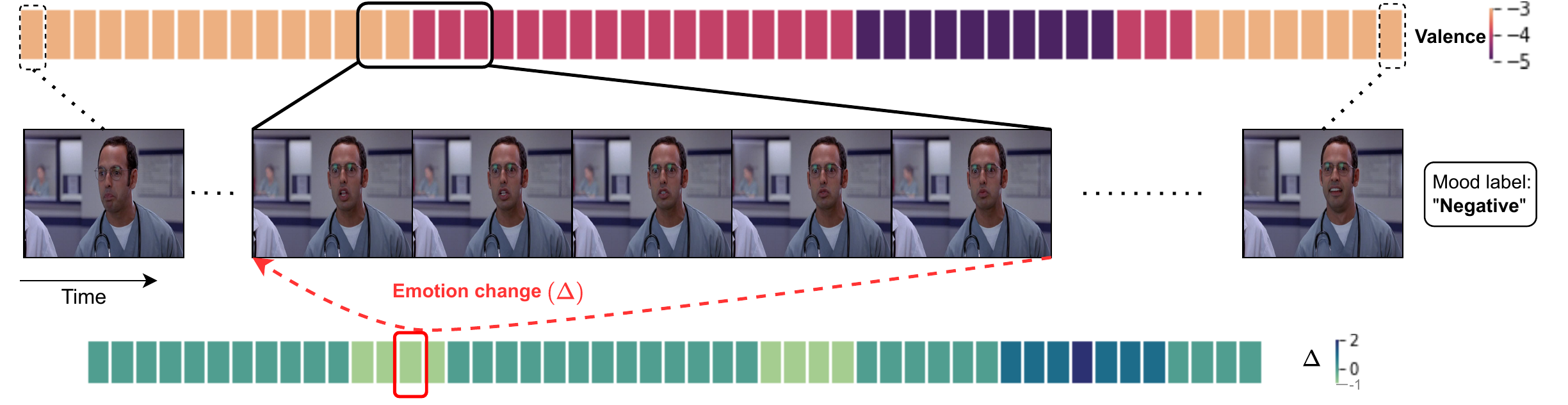}\vspace{-2mm}
\caption{\textbf{Problem Illustration:} Figure depicts emotion changes in an input video sample having a negative mood label. The top colour bar denotes per-frame valence values for the video, while the bottom colour bar depicts emotional valence change ($\Delta$) labels over a window of five frames (best viewed in colour).}\vspace{-2mm}
\label{fig:mood_delta_overview}
\end{teaserfigure}

\maketitle

%
%

\section{Introduction}
\label{sec:introduction}

\begin{figure*}[t]
\centering
\includegraphics[height=3.5cm, width=0.4\textwidth]{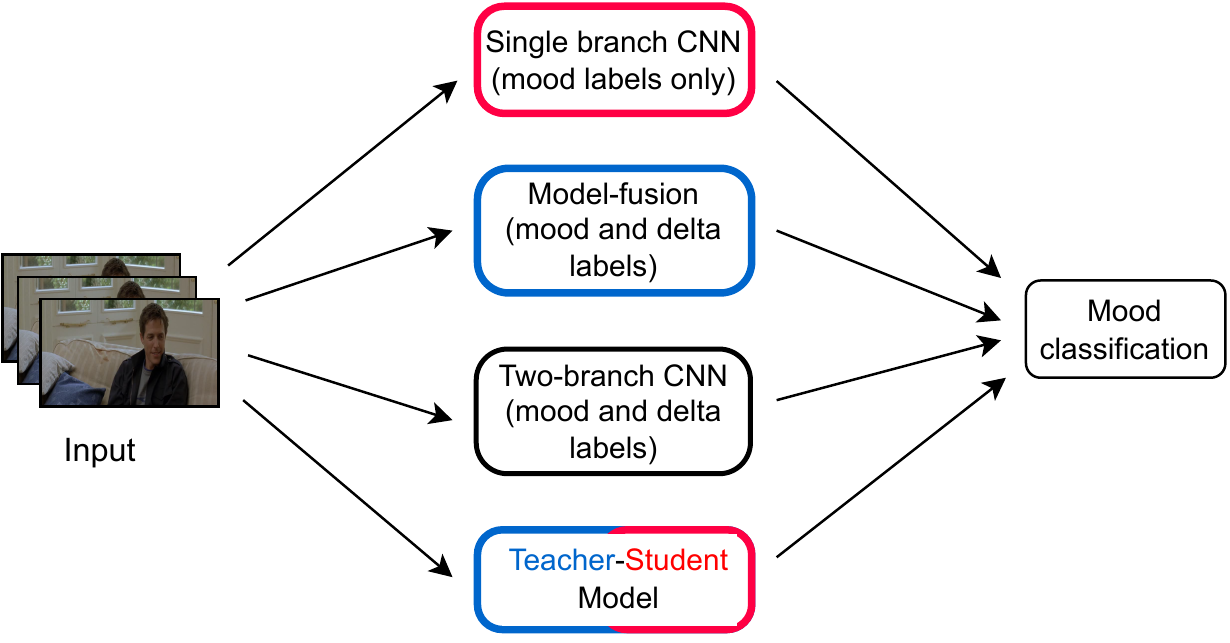}\vspace{-2mm}
\caption{Overview of the proposed mood recognition framework.}\vspace{-3mm}
\label{fig:overview}
\end{figure*}

There is mounting evidence that emotions play an essential role in rational and intelligent behaviour. Besides contributing to a richer quality of interaction, they directly impact a person's ability to interact in an intelligent way \cite{picard2000affective}. Quite often, the terms \emph{mood} and \emph{emotion} are used interchangeably, although differences in duration, intensity and attribution exist. While \textbf{\emph{emotion}} is a short-term affective state induced by a source which can last for a few minutes, \textbf{\emph{mood}} refers to a longer-term affective state that can last for hours or even days, and be without a causal source~\cite{jenkins1998human}. Most research in affective computing has focused on inferring emotional states, while very little research has so far been devoted to automated mood recognition~\cite{katsimerou2015predicting} or the joint modelling of the interplay between emotion and mood for improved affective state recognition. 
Psychological studies on mood have made substantial progress. An eye-tracking study has revealed that positive mood results in better global information processing than a negative mood \cite{schmid2011mood}. The authors in~\cite{schmid2010mood} have observed a mood-congruity effect, where positive mood hampers the recognition of mood-incongruent negative emotions and vice-versa. The mood-emotion loop is a theory that posits mood and emotion as distinct mechanisms, which affect each other repeatedly and continuously. This theory argues that mood is a high-level construct activating latent low-level states such as emotions~\cite{wong2016mood}. Recognising the interactions between mood and emotion has the potential to lead to a better understanding of affective phenomena, such as mood disorders and emotional regulation. 

On the contrary, mood recognition has rarely been addressed from a computational perspective and only a few studies have explored mood~\cite{katsimerou2015predicting}. Body posture and movement correlates of mood have been explored in~\cite{thrasher2011mood}. User mood is induced via musical stimuli and the authors have observed that head posture and movements characterise happy and sad mood.
Katsimerou \emph{et al.}\ \cite{katsimerou2015predicting} have examined automatic mood prediction from recognised emotions, showing that clustered emotions in the valence-arousal space predict single moods much better than multiple moods within a video. Research on mood prediction has also neglected to investigate the interplay between mood and emotion, though the psychological literature recognises a relationship between the two \cite{morris1992functional}.

From an affective computing viewpoint, developing a mood recognition framework requires ground-truth mood labels for model training, but only very few databases record the user mood (directly or indirectly via an observer). Widely used affective corpora, such as AFEW-VA~\cite{kossaifi2017afew}, HUMAINE~\cite{douglas2011humaine}, SEMAINE~\cite{mckeown2011semaine} and DECAF~\cite{DECAF}
only contain dimensional and/or categorical emotion labels. One of the few datasets with mood ratings is EMMA~\cite{katsimerou2016crowdsourcing}, where the annotations developed represent the overall emotional impression of the human annotator (or observer) for the examined stimulus~\cite{katsimerou2016crowdsourcing}. Machine learning approaches have been extensively used for inferring emotions from visual, acoustic, textual and neurophysiological data~\cite{shukla22, bilalpur17, liu2018multi, calix2010emotion, trigeorgis2016adieu}. Contemporary studies emphasise the improved performance of multimodal approaches to the detection of emotional states vis-\'a-vis unimodal ones~\cite{d2012consistent}. Recent studies characterise mood disorders, such as depression, by examining speech style, eye activity, and head pose~\cite{alghowinem2016multimodal, senoussaoui2014model, alghowinem2013joyous}. Deng \emph{et al.}\ \cite{deng2020multitask} propose a multitask emotion recognition framework that can deal with missing labels employing a teacher-student paradigm. \emph{Knowledge Distillation} (KD) is a technique that enables the transfer of knowledge between two neural networks, unifying model compression and learning with privileged information~\cite{hinton2015distilling, lopez2015unifying}. KD techniques have been employed for facial expression recognition where the teacher has access to a fully visible face, whereas the student model only has access to occluded faces \cite{georgescu2021teacher}.
%

While our research is ultimately aimed towards mood prediction and understanding the interplay between mood and emotions from video data, the present study is an initial step on this path. We use the AFEW-VA dataset to derive (a) \emph{dominant emotion} labels, which refer to the emotion persisting for most consecutive frames (termed \textit{mood} labels), and (b) $\Delta$ or emotion change labels, which represent the change in emotion over a fixed window size. Given the sparsity of in-the-wild data with mood annotations and the preliminary nature of this study, the dominant emotion labels are used here in lieu of actual mood labels. In the future, we will be using actual mood labels derived from expert annotators.
Fig.~\ref{fig:mood_delta_overview} illustrates how emotion change is captured for an exemplar video clip, while Fig.~\ref{fig:overview} overviews our dominant emotion or mood prediction framework. A unimodal 3D Convolutional Network Network (3D CNN) is trained using only mood labels, while a two-branch (multimodal) CNN model, multi-layer perceptron, and a teacher-student model are evaluated for fusing emotion-mood information for mood prediction. Empirical evaluation reveals that incorporating emotion change information improves mood prediction performance by as much as 54\%, confirming the salience of fine-grained emotional information for coarse-grained mood prediction. This study makes the following contributions: 
\begin{itemize}
    \item To the best of our knowledge, from a computational modelling perspective, this is the first study to examine mood prediction incorporating both mood and emotional information. Mood labels are derived from valence annotations, instead of subjective impressions provided by a human annotator.
    \item The experimental evaluation of multiple models shows that incorporating emotional change information is beneficial and can produce a significant improvement in mood prediction performance.
\end{itemize}


%
%

\section{Materials} \label{sec:materials}

\begin{figure*}[t]
\centering
\includegraphics[height=3.5cm, width=0.8\linewidth]{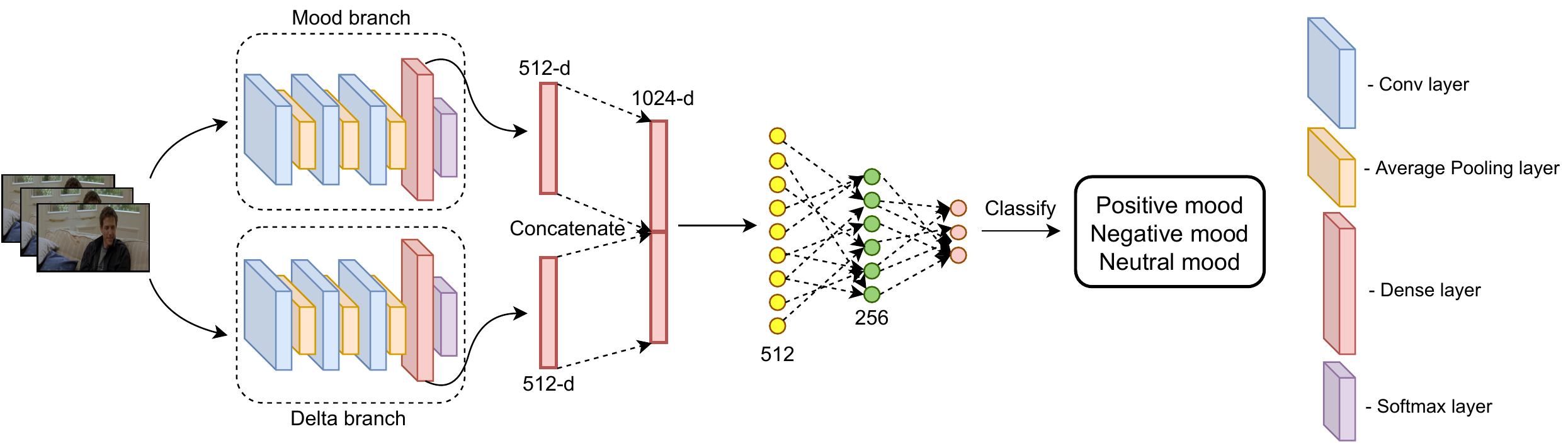}
\caption{Architecture of the model fusing mood and $\Delta$ information. The legend on the right side shows the unique colours used to describe the layers in the architecture. (Best viewed in colour)}\vspace{-3mm}
\label{fig:2-CNN+MLP}
\end{figure*}

\subsection{Dataset} \label{sec:dataset}

Here, the AFEW-VA~\cite{kossaifi2017afew} dataset, a subset of the AFEW~\cite{dhall_AFEW}, comprising 600 video clips extracted from feature films at a rate of 25 frames per second, was used. 
Video clips in this dataset range from very short sequences ($\sim$ 10 frames) to longer videos (145 frames), and depict various facial expressions. The videos are captured in both indoor and outdoor naturalistic settings~\cite{dhall_AFEW}. There are 240 subjects in AFEW-VA, who are the actors in the videos. Each clip has per-frame valence and arousal annotations in the $[-10,10]$ range, annotated by two expert annotators (1 male, 1 female). To examine the interplay between mood and emotion, and perform mood prediction, we assigned a mood label to each clip.

%
%

\subsection{Labels} \label{sec:labels}

\subsubsection{\textbf{Mood labels:}} The valence value prevailing over most consecutive frames
is considered as the dominant emotion label for a video. As mentioned before, given the lack of in-the-wild datasets with both mood and emotion labels and given the preliminary nature of this study, we consider the dominant emotion label as the mood label in this paper to assess the proposed framework (see Sec.\ \ref{sec:introduction}).
We assigned the valence ranges of $\left[-10, -3\right)$, $\left[-3, 3\right]$ and $\left(3, 10\right]$, to labels of -1 (negative), 0 (neutral), and 1 (positive), respectively. Based on this annotation scheme, the AFEW-VA dataset comprises 59, 400, and 141 videos with positive, neutral, and negative mood labels, respectively. We consider an overlapping sequence of $k$ frames as one input sample. For example, considering a video clip with $n=10$ frames and $k=3$, the input samples include frames [1--3], [2--4], [3--5], $\ldots$, [8--10]. With $k = 5$, the data comprises a total of 27,651 samples. Each sample is assigned the mood label of its source clip. 

\subsubsection{\textbf{Emotion Change ($\Delta$) labels:}} Apart from the mood label, we also associate a $\Delta$ to every sample, which denotes the change in emotional valence over $k$ frames. For a video clip with $n$ frames, $\Delta$ is computed as the valence difference between the $t^{th}$ and $(t-k+1)^{th}$ frames for $t = k, k+1, ...,n$. \textit{E.g.}, considering $k=5$, if $v_5 = -5$ and $v_1 = -3$, where $v_5$ and $v_1$ denote the valence of the fifth and the first frame, respectively, $\Delta = v_5 - v_1 = -2$. As a first step towards mood prediction, we assign the \emph{sign} of the $\Delta$ value to be the $\Delta$ label ($\Delta$ = -1 in the previous example). $\Delta$ label is computed as:
\begin{equation}
\sgn \Delta =\begin{cases} 
-1 & \text{if } \Delta < 0, \\
0 & \text{if } \Delta = 0, \\
1 & \text{if } \Delta > 0. \end{cases}
\end{equation}
Thus, each input video sample has both mood and $\Delta$ labels $\in [-1, 0, 1]$. 

%
%
\section{Methods} 

\label{sec:methods}
Our work leverages the interplay and mutual influence between mood and emotion~\cite{katsimerou2015predicting, wong2016mood}. We utilise valence annotations to gather information on mood, and perform mood classification using Convolutional Neural Networks. This section describes the unimodal (1-CNN) and multimodal (2-CNN, 2-CNN+MLP, and Teacher-Student (TS) network) models, their architectures and the hyper-parameters employed for model training. 

%
%

\subsection{Single-Branch Mood Classification} \label{sec:single-mood}

For mood classification, we feed a single-branch three-layered CNN (or 1-CNN) with input video samples and mood labels as described in Sec.\ \ref{sec:dataset}. The 1-CNN is denoted using a dashed-rectangle in Fig.\ \ref{fig:2-CNN+MLP}. Each convolutional layer convolves the input sample with a stride of 3, and the three convolutional layers comprise 16, 32 and 32 kernels of size $3 \times 3 \times 3$, respectively. Each of these layers is followed by average pooling over 2-pixel regions. The output of the third convolutional layer is flattened, followed by batch normalisation. The dense layer comprises 512 neurons, followed by a SoftMax layer with three neurons corresponding to the three mood classes considered in this study.

The input dimensionality for the 3D-CNN is $5 \times 32 \times 32 \times 3$, with each input sample comprising five frames of size $32 \times 32 \times 3$. We use categorical cross-entropy as the loss function for training the model. The fine-tuned hyper-parameters include a learning rate $\in \{10^{-5}, 10^{-3}\}$, batch size $\in \{64, 128, 256\}$, and dropout rate $\in \{0.4, 0.5\}$. The Adam optimizer is used for optimising model learning with stochastic gradient descent.

%
%
\subsection{Model Fusion} 
\label{sec:model_fusion}
\begin{figure*}[t]
\centering
\includegraphics[height=3.5cm, width=0.8\linewidth]{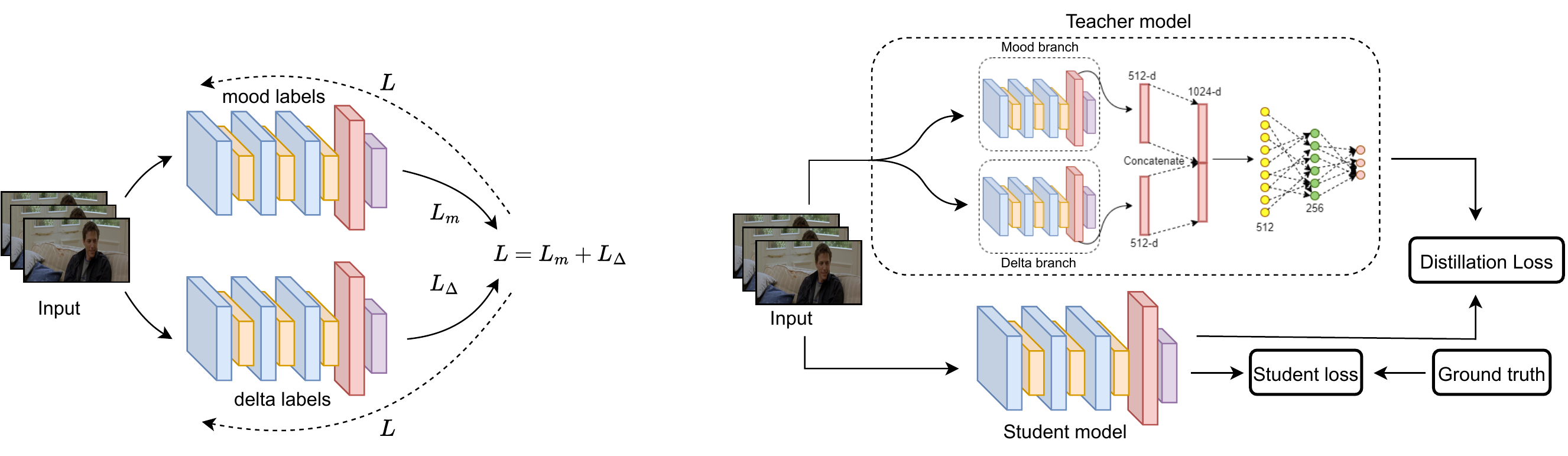}
\caption{\textbf{(Left)} Architecture of the 2-CNN model, composed of two 1-CNN models. \textbf{(Right)} Architecture of the TS-Network. The layers of the model are described in Fig.\ \ref{fig:2-CNN+MLP}. (Best viewed in colour)}.\vspace{-3mm}
\label{fig:2-CNN_and_TS-Network}
\end{figure*}

Given the success of fusion-based emotion inferring approaches \cite{liu2018multi, alghowinem2016multimodal}, we fused the mood and $\Delta$ information for mood prediction. To examine the influence of emotion change or $\Delta$ on mood, we employ a two-branch CNN model with a Multi-Layer Perceptron (2-CNN+MLP). As shown in Fig.\ \ref{fig:2-CNN+MLP}, the two three-layered 1-CNNs in each branch are trained independently, with $\Delta$ labels fed to one branch and mood labels to the other. From the trained models in each branch, we gather a 512-dimensional vector from the penultimate CNN layer. The vectors from the two branches are concatenated to form a 1024-dimensional feature, which is then passed to an MLP for mood classification. This architecture assumes the availability of $\Delta$ labels in both the training and test phases. 

The two branches involve identical 1-CNN networks and only differ with respect to their input labels. The input dimensionality and the hyper-parameters for each branch are identical to 1-CNN.
The MLP has two dense layers with 512 and 256 neurons, respectively, and a SoftMax layer with three neurons to classify the mood. Overall, we fuse representations learned in the two branches and feed them to an MLP for mood prediction.

%
%
\subsection{Two-Branch Mood Classification}
%

An alternative to the feature fusion model is to perform {end-to-end} optimisation \cite{trigeorgis2016adieu, graves2014towards} with the mood and $\Delta$ branch networks. To this end, we employ a 2-CNN model (Fig.\ \ref{fig:2-CNN_and_TS-Network} (left)), a two-branch model with end-to-end learning, with a three-layered 1-CNN model in each branch. As for the fusion model, the 2-CNN model is composed of the mood and $\Delta$ 1-CNN branches. However, this model differs from 2-CNN+MLP with respect to the training process. The categorical cross-entropy losses from the mood and $\Delta$ branches ($L_m$ and $L_\Delta$) are summed up, and the cumulative loss is minimised in this model. 

Unlike 2-CNN+MLP, which requires $\Delta$ labels in the test phase, $\Delta$ labels serve as auxiliary information and are only incorporated during 2-CNN model training and back-propagation. Only mood labels are utilised during the test phase, as the mood branch alone is activated for inference. 

%
%

\subsection{Teacher-Student Network}
In addition to model fusion and end-to-end optimisation, we employ knowledge distillation~\cite{hinton2015distilling} to transfer knowledge from a larger teacher network to a smaller student network. Fig.\ \ref{fig:2-CNN_and_TS-Network} (right) presents the Teacher-Student Network, where the teacher and student networks are as described in Sec.\ \ref{sec:model_fusion} and \ref{sec:single-mood}, respectively. The pre-trained teacher network utilizing both mood and $\Delta$ labels (2-CNN+MLP) distills knowledge, while training the student network (1-CNN) only requires mood labels. As inference is again based on the student network, $\Delta$ labels are not required during test time.  

The student model's SoftMax layer involves a hyper-parameter called the \emph{temperature} $T$, which controls the smoothness of the output probabilities. Setting $T > 1$ increases the weight of smaller logit\footnote{The logit function is the quantile function associated with the standard logistic distribution.} values, thus revealing more information about inter-class relations than the one-hot labels \cite{hinton2015distilling}. The Kullback–Leibler (KL) divergence is used to compute the distillation loss, while sparse categorical cross-entropy is used as the student loss function. The overall loss of the teacher-student model $L_{TS}$ is the weighted sum of the student loss $L_{\text{stu}}$ and distillation loss $L_{\text{dis}}$:
\begin{equation}
L_{TS} = \alpha L_{\text{stu}}  + (1 - \alpha) L_{\text{dis}}
\end{equation}
\noindent where $\alpha$ is a training hyper-parameter. The fine-tuned hyper-parameters include a batch size $\in \{16, 64, 128\}$, $T$ $\in \{3, 5, 7\}$ and $\alpha \in \{0.05, 0.1, 0.15, 0.2, 0.25, 0.3\}$.  

%
%

\subsection{Performance Measures}
All models are evaluated via subject-independent 5-fold cross-validation to ensure the same subject does not appear simultaneously in two different folds. We report the mean accuracy over the five folds as the metric for performance evaluation.

%
%

\section{Results and Discussion} \label{sec:results}
\begin{table*}[t]
\caption{Performance comparison of different models with a 5-frame input sequence ($k = 5$).}
\label{tab:results_main}
\resizebox{\textwidth}{!}{%
\begin{tabular}{@{}c|cc|c|c|cc@{}}
\textbf{Model} & \multicolumn{2}{c|}{\textbf{1-CNN}} & \textbf{2-CNN} & \textbf{2-CNN + MLP} & \multicolumn{2}{c}{\textbf{TS-Network}} \\ 
\hline
\textbf{Label} & \textbf{Mood} & \multicolumn{1}{c|}{\textbf{Delta}} & \textbf{Mood} & \multicolumn{1}{c|}{\textbf{Mood}} & \textbf{\begin{tabular}[c]{@{}c@{}}Mood\\  (student without teacher)\end{tabular}} & \textbf{\begin{tabular}[c]{@{}c@{}}Mood\\ (student with teacher)\end{tabular}} \\ 
\hline
Accuracy ($\mu \pm \sigma$) & \multicolumn{1}{l}{0.35 $\pm$ 0.10} & 0.53 $\pm$ 0.11 & \multicolumn{1}{l|}{0.73 $\pm$ 0.06} & 0.87 $\pm$ 0.15 & 0.35 $\pm$ 0.10 & \textbf{0.89 $\pm$ 0.09} \\ \hline
\end{tabular}%
} \hspace{-2mm}
\end{table*}

\begin{figure*}[t]
\centering
\includegraphics[height=4.5cm, width=0.5\textwidth]{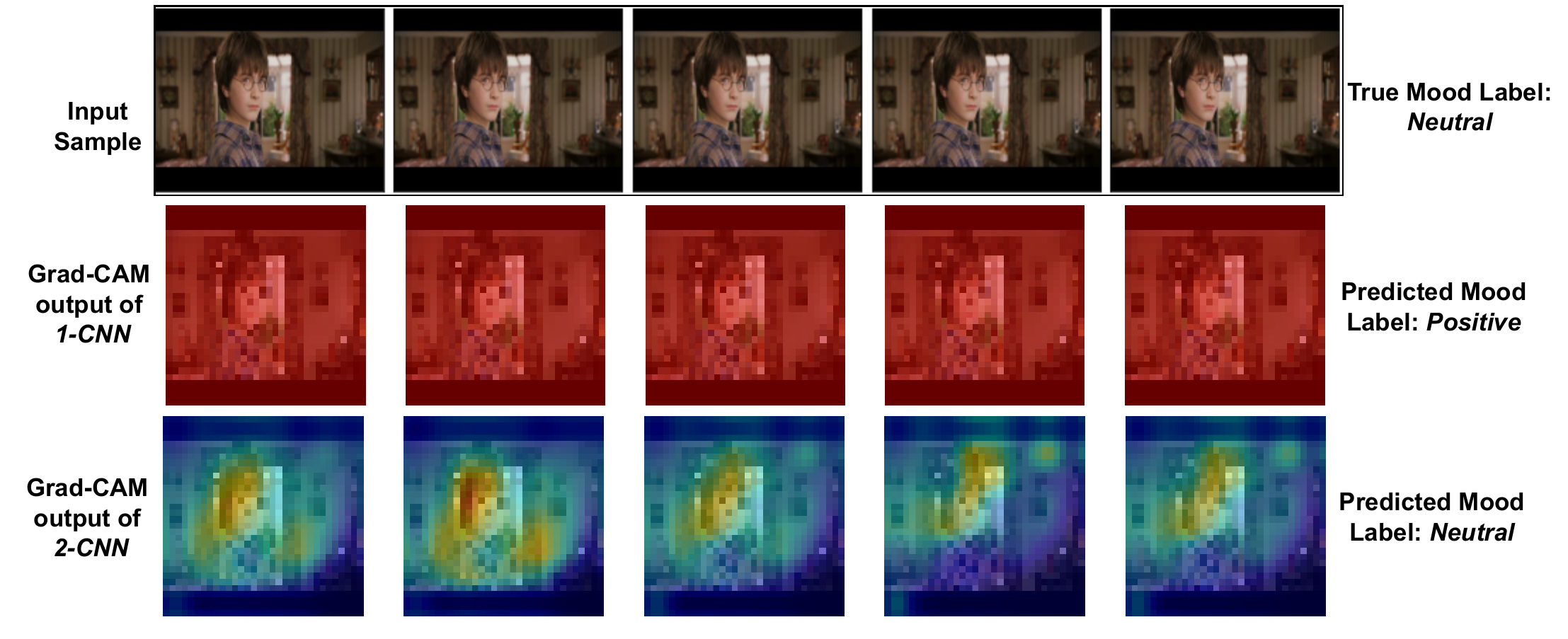}
\caption{GradCAM maps~\cite{selvaraju2017grad} depicting improved mood prediction when $\Delta$ is learnt by focusing on relevant face parts. \emph{(Top)} An input sample with the ground truth mood label being neutral. \emph{(Centre)} GradCAM maps of the 1-CNN model, with the predicted label as positive. \emph{(Bottom)} Correct prediction using the 2-CNN model. (Best viewed in colour).}\vspace{-2mm}
\label{fig:grad_cam}
\end{figure*}

Table \ref{tab:results_main} shows the results obtained for each of the models implemented. The 1-CNN model is trained independently with mood and $\Delta$ labels. The model trained with $\Delta$ labels yields a higher accuracy as compared to the model trained with mood labels, implying that the 1-CNN model is able to learn temporal emotional changes better than the high-level mood construct. The 2-CNN+MLP model, combining information from the mood and $\Delta$ CNN branches for mood prediction, achieves a much higher accuracy than the 1-CNN model. This result validates our hypothesis that incorporating emotion change ($\Delta$) information improves mood prediction. Further, the 2-CNN model, with its independent $\Delta$ and mood branches, performs better than the 1-CNN model but worse than the 2-CNN model, again confirming that fusing mood and $\Delta$ information is beneficial. Finally, the TS-Network composed of the 2-CNN+MLP as the teacher model trained with mood and auxiliary $\Delta$ labels, and the student 1-CNN network trained with mood labels gives the highest accuracy among all models. Cumulatively, these results confirm that modelling emotional change information is beneficial for mood prediction. 

Fig.\ \ref{fig:grad_cam} qualitatively illustrates how the incorporation of $\Delta$ information induces better attention from the classifier. The 1-CNN model trained using only mood labels is unable to focus on the face and bases its incorrect prediction on both the facial and background information. In contrast, the 2-CNN model trained with both mood and $\Delta$ labels is able to focus on the face to make the correct mood prediction, consequently improving mood prediction performance.  

As $\Delta$ is computed over a fixed window size, it captures emotion change over a shorter duration than the actual video. Incorporating $\Delta$ information for mood classification implies that the focus is not just on the person's mood, which lingers on for a longer duration, but also on the shifts and variations in short-term emotions that are inherent. This aligns with real-world scenarios where a person in a particular mood could experience emotional fluctuations under different circumstances. For example, while in a negative mood, a person could experience a positive emotion following exposure to a positive stimulus, or in a positive mood, one could experience a negative emotion following a negative event. Here, the interplay between emotion and mood would have a dampening effect. The emotion need not necessarily be of opposite valence; an emotion of the same valence as the mood could create a variation (amplification) as well. With the experimental results obtained, it is noteworthy that apart from mood, which speaks of the holistic affective state (as it lingers in the background of one's consciousness for a long duration), the local affective state captured as emotion change is positively contributing to mood prediction, which supports our hypothesis that emotion and mood should be modelled simultaneously for improved affective state prediction. 

%
%

\section{Conclusions} \label{sec:conclusions}
Increasing evidence in psychology shows a close association between the affective states, mood and emotion, while little has been done in this regard from a computational modelling perspective. While our long-term aim is to examine mood and investigate the interplay between mood and emotions, as a first step, we use the AFEW-VA database, which has per-frame annotations of valence and arousal, and derive the dominant emotion labels from the valence values as an approximation of the mood labels. In addition to these labels, we explore the potential of temporal emotion change for mood prediction. A unimodal CNN and multimodal feature fusion (2-CNN+MLP), 2-CNN and TS networks have been explored in this study. The experimental results demonstrate that learning the emotion change greatly improves mood prediction. 

The present study is limited with respect to finding the existence of a relation between the two affective states. The work is also limited in considering the window size for the input sample. In the future, we will investigate the interplay between mood and emotions, by considering actual mood labels derived from expert annotators. We also plan to examine the mutual influence of emotion on mood by taking into account the polarity of the affective states. Exploring the effect of different window sizes for capturing emotion change is also left to future work. 

%
%

\begin{acks}
This research was supported partially by the Australian Government through the Australian Research Council's Discovery Projects funding scheme (project DP190101294).
\end{acks}


\bibliographystyle{ACM-Reference-Format}
\bibliography{references}
\thispagestyle{plain}
\end{document}